\documentclass[sigconf,hyphens,nonacm]{acmart}

\usepackage{multirow}
\usepackage{tabularx}
\usepackage{graphicx}
\usepackage{listings}
\lstdefinestyle{requesttrace}{
  basicstyle=\ttfamily\scriptsize,
  breaklines=true,
  columns=fullflexible,
  frame=single
}
\AtBeginDocument{%
  }

\makeatletter
\AtBeginDocument{\renewcommand{\@copyrightpermission}{}}
\AtBeginDocument{%
  \fancypagestyle{firstpagestyle}{%
    \fancyhf{}%
    \fancyfoot[C]{\thepage}%
  }%
}
\makeatother
\AtBeginDocument{\pagestyle{plain}}

\begin{document}

\title[Developer Experience with AI Coding Agents]{Developer Experience with AI
  Coding Agents: HTTP Behavioral Signatures in Documentation Portals}
\titlenote{This is the extended version of the work. A short version
  (extended abstract) appeared as: Oleksii Borysenko. 2026.
  \emph{Documentation Retrieval by AI Coding Agents: An Empirical Study of HTTP
  Behavioral Signatures}. In \emph{Proceedings of the 1st International
  Workshop on Agentic AI for Next-Generation Software Development
  (AgenticDev~'26), co-located with ASE~2026}, Munich, Germany. ACM,
  pp.~168--169. \url{https://doi.org/10.1145/3843282.3844445}.
  When citing this work, please cite the peer-reviewed ACM version; cite this
  arXiv version (arXiv:2604.02544) when referring to material available only
  here (see ``How to Cite'' at the end of the paper).
  arXiv subject classification: cs.SE (Software Engineering).}

\author{Oleksii Borysenko}
\email{borysenko@cisco.com}
\orcid{0009-0000-2965-7661}
\affiliation{%
  \institution{Cisco DevNet}
  \country{}
}

\renewcommand{\shortauthors}{Borysenko}

\begin{abstract}
The rapid adoption of AI coding agents and AI assistant web services is
fundamentally changing how developers discover, consume, and interact with
technical documentation. This paper studies that transformation across three
interconnected dimensions: documentation accessibility, content analytics, and
feedback systems. We present an empirical study of HTTP request fingerprints
from nine AI coding agents (Aider, Antigravity, Claude Code, Cline, Cursor,
Junie, OpenCode, GitHub Copilot (VS~Code agent mode), and Windsurf) and six AI assistant services (ChatGPT, Claude,
Google Gemini, Google NotebookLM, MistralAI, and Perplexity) accessing a live developer
documentation endpoint, revealing identifiable behavioral signatures in HTTP
runtime environments, pre-fetch strategies, User-Agent strings, and header
patterns. Our study shows that AI agent access compresses multi-page
navigation into a single or two requests, making traditional engagement
metrics---session depth, time-on-page, click path, and bounce rate---unreliable
indicators of actual documentation consumption. We discuss practical adaptations
for developer portal teams, including tokenomics-aware documentation design,
adoption of emerging machine-readable standards (\texttt{AGENTS.md},
\texttt{llms.txt}, \texttt{skill.md},
\texttt{agent-\allowbreak{}permissions.json}), MCP server--based feedback
channels, and analytics instrumentation for AI referral traffic.
\end{abstract}

\ccsdesc[500]{Software and its engineering~Software maintenance tools}

\keywords{%
  developer experience, AI coding agents, documentation accessibility,
  content analytics, feedback systems, HTTP fingerprinting,
  developer portals, tokenomics%
}

\maketitle
\thispagestyle{plain}


\section{Introduction}

Developers gain experience working with documentation, APIs, SDKs, and MCP
servers. For many developers, interaction with developer-oriented documentation
now primarily occurs within AI coding agents and IDEs with AI features. AI
coding agents use tools such as web search that allow them to stay within the
IDE perimeter alongside other coding assistants. Developer experience has
shifted significantly: users now spend more time with AI assistants and
coding agents---according to the Stack Overflow Developer Survey, 51\% of
professional developers use AI tools
daily~\cite{stackoverflow2025}. As a result, developer attention that was previously directed toward
documentation and code examples is increasingly mediated through the interfaces
of AI assistants.

Documentation for open-source or proprietary products can be accessible via
AI tools as part of an MCP server or as part of model training datasets.
As a general rule, publicly accessible pages on sites that don't block AI
crawlers in their \texttt{robots.txt} files are crawled by AI agent crawlers
and incorporated into model training data. The problem is that
this data can become outdated rapidly---for example, when a company releases a
new API version or deprecates endpoints. Updates are not automatically
propagated to related LLMs or AI assistants; some update cycles take months.
Each model has a knowledge cut-off date. Cheng et al.~\cite{cheng2024}
demonstrate that LLMs' effective knowledge cut-offs often diverge significantly
from their reported training dates, meaning documentation ingested during
training may be far more outdated than vendors claim. This is why AI assistants
and agents also rely on retrieval tools such as web search to obtain the latest
information directly from source websites. Open-source projects like
Context7~\cite{context7} help pull up-to-date, version-specific documentation
and code examples straight from the source.

Writing code is one of the primary niches where Generative AI providers and AI
coding agent companies address enterprise needs. Companies still seek to quantify
efficiency and return on investment, since developers' time spent guiding
coding agents incurs a cost. Brandebusemeyer et al.~\cite{brandebusemeyer2026}
found in an empirical field study at SAP that moderate use of GenAI improves
efficiency and reduces perceived workload, while excessive or combined use of
multiple interaction types diminishes these benefits.

Developer Experience (DevX) is now being formally recognized as a research
discipline. Combemale~\cite{combemale2025} argues that DevX ``profoundly
influences critical development activities and overall productivity, especially
as development becomes increasingly collaborative and diverse in terms of
application domains.'' This paper explores how these shifts manifest
specifically in the context of AI coding agents, where documentation
accessibility, content analytics, and feedback systems may require fundamental
rethinking.

AI bot activity has surged. Cloudflare~\cite{cloudflare2024} measured the volume
of requests from AI crawlers, identifying Bytespider, Amazonbot, ClaudeBot, and
GPTBot as the highest-volume sources. Bilal et al.~\cite{bilal2025} cite an
18\% year-over-year increase in non-human web traffic, with GPTBot requests
alone growing more than 300\% as AI systems progressively replace human web
consumption.

The primary objective of this study is to characterise how current AI coding
agents and AI assistant web services retrieve developer documentation through
HTTP, and to examine the implications of these retrieval patterns for
developer-portal analytics.

\section{Developer Documentation in the AI Agent Era}

Portals and websites that host developer and engineering information are likely
to experience significant declines in organic search traffic~\cite{kuryatnik2025} while
seeing increased visits from AI crawlers~\cite{liu2025} and bots. Users will
consume content on the platforms they are accustomed to---coding agents, AI
assistants, and IDEs. Some SaaS documentation platforms and hosted portals are
adding features to ensure every developer-oriented page is discoverable by both
humans and AI alike: content is automatically indexed by search engines, and AI
agents receive pages in clean Markdown rather than raw HTML, making them faster
to process and more token-efficient. Developers can view the Markdown source of
any page themselves by appending \texttt{.md} to its URL~\cite{mintlify2024}.
If the content covers API or MCP documentation, it should be easily accessible
for copying or downloading via a single stable link.

\subsection{Platforms Where Developers Consume Content}

There are many platforms where users can now consume developer-oriented content
outside of IDEs and coding agents---developer guides, demo use cases, technology
overviews, and tool introductions. Users select the platforms they prefer. For
example, a developer might summarize text, convert a developer-oriented blog
post into audio using NotebookLM~\cite{tufino2025}, or convert audio and video
into searchable, interactive transcripts. Developers choose different modalities
for consuming developer-oriented content.

\begin{figure}[h]
  \centering
  \includegraphics[width=\linewidth]{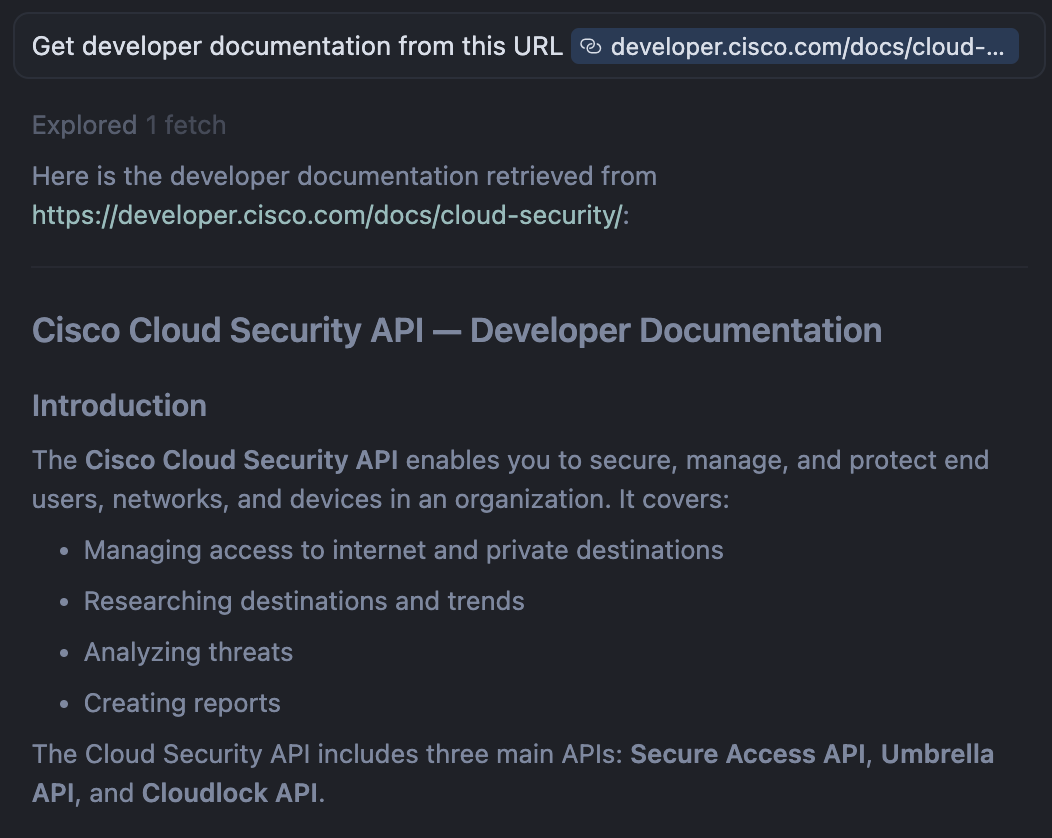}
  \caption{Cursor AI coding agent retrieving developer documentation from a
    developer portal in a single fetch. The prompt \emph{``Get developer
    documentation from this URL''} triggers one HTTP request that pulls the
    entire page, replacing multi-step human navigation with a single
    machine-readable response.}
  \label{fig:cursor-chat}
\end{figure}

An important consideration for developer portal analytics and maintainers is
that when users copy or consume content on an external platform, they are more
likely to continue interacting with that external platform and to ask follow-up
questions or report issues there. In this case, critical portal analytics
data---time on page, UI interaction events, and link clicks---becomes
unavailable. Gathering developers' comments and feedback becomes correspondingly
harder. An example of this single-fetch retrieval is shown in Figure~\ref{fig:cursor-chat}.

\section{Research Method}
\label{sec:method}

This study uses a controlled comparative design to examine how AI coding agents
and AI assistant web services retrieve developer documentation. We focus on
observable HTTP-level behavior: request method and sequence, HTTP runtime
signals, User-Agent strings, selected request headers, and discovery-file
access. The goal is to identify reproducible behavioral signatures rather
than to estimate the prevalence of particular behaviors across all AI tools.

\subsection{Research Questions}

The study addresses the following research questions:
\begin{itemize}
  \item \textbf{RQ1:} What HTTP request signatures distinguish the evaluated
        AI coding agents and AI assistant web services?
  \item \textbf{RQ2:} How stable are the observed HTTP request signatures
        across repeated trials and network conditions?
  \item \textbf{RQ3:} What implications do the observed HTTP retrieval patterns
        of AI agents have for developer-portal analytics based on multi-page,
        client-side interaction?
\end{itemize}

RQ1 is addressed through the request-fingerprint analysis in
Section~\ref{sec:fingerprints}; RQ2 through the repeated-trial comparison
described in Section~\ref{sec:method}; and RQ3 through the implications for
developer-portal analytics discussed in Section~\ref{sec:analytics}.

\subsection{Tool Selection and Scope}

We selected nine AI coding agents and six AI assistant web services using a
purposive sampling strategy. The coding-agent sample prioritised developer-facing
tools with practical availability for repeatable testing and diversity in product
form, including command-line agents, AI-native IDEs, IDE-integrated assistants,
and browser-based services. The sample includes prominent tools identified in the
broader AI developer-tools ecosystem. In particular, the SlashData \emph{AI Coding
Tools Benchmark} benchmarks 20 AI coding assistants, agents, and AI-native IDEs
using responses from 2,393 professional developers across adoption, satisfaction,
workflow reliance, task quality, trustworthiness, and perceived productivity
gains~\cite{slashdata2026}.

The sample was limited to tools that were open source, freely available, or
accessible through a free or trial tier during the measurement period. We
excluded enterprise-only tools, services requiring unavailable organisational
credentials, and tools that could not be tested directly and repeatably.
Accordingly, the study characterises the observed tools and does not claim
statistical representativeness of all AI coding products.

\subsection{Endpoint and Experimental Environment}

Measurements used a purpose-built, publicly accessible developer documentation
endpoint implemented with Node.js and Express. The endpoint served
developer-oriented content and API documentation and exposed \texttt{robots.txt}
(permitting crawlers) and \texttt{llms.txt}. This controlled environment allowed
the same URL, content, and discovery files to be presented to every tested tool.
The endpoint did not require authentication and did not rely on client-side
rendering.

The \texttt{robots.txt} file permitted all user agents to access the portal
except the \texttt{/guide} path, and specified the portal sitemap. The
\texttt{llms.txt} file identified the portal as an API documentation portal and
linked to its main developer-documentation page and OpenAPI 3.0.1
specification. Neither discovery file changed during the measurement period.
Their contents were:

\begin{lstlisting}[style=requesttrace]
User-agent: *
Allow: /
Disallow: /guide

Sitemap: https://[second-level domain].dev/sitemap.xml
\end{lstlisting}

\begin{lstlisting}[style=requesttrace]
# [second-level domain].dev

> API documentation portal

## Documentation

- [Developer documentation and API](https://[second-level domain].dev/): Main API documentation page with endpoints, rate limits, and request headers
- [OpenAPI Specification](https://[second-level domain].dev/openapi.yaml): OpenAPI 3.0.1 specification for the API
\end{lstlisting}

\subsection{Trial Protocol and Data Collection}

Each tool received the identical prompt:
\emph{``Get developer documentation from this URL (\ldots{})''.}
AI coding agents were tested through their normal user-facing workflow, while
AI assistant services were tested by submitting the same prompt and URL through
their chat interfaces. Each tool was tested in three independent trials during
February--March 2026. Trials were conducted from the same local network and
from external networks.

All inbound requests were recorded exclusively by server-side Express
middleware. For each request, the middleware captured the HTTP method and
version, requested URL, complete header set, User-Agent string, client IP
address. No client-side JavaScript instrumentation was used; the collected data
therefore represent raw HTTP traffic reaching the endpoint.

\subsection{Analysis and Consistency Assessment}

We compared request traces across trials for each tool, examining request count
and ordering, access to \texttt{/robots.txt} and related discovery paths, HTTP
method, User-Agent, \texttt{Accept} headers, \texttt{Sec-Fetch-*} headers, and
other stable runtime signals. A fingerprint was considered consistent when the
core request sequence and header pattern were unchanged across the three trials.
No systematic differences in the reported signatures were observed between local
and external-network trials.

\subsection{Scope and Threats to Validity}

The use of one controlled documentation endpoint strengthens comparability but
limits external validity. Agent behavior may differ for static HTML portals,
client-side-rendered sites, authenticated documentation, rate-limited services,
or websites using anti-bot protections. The results should therefore be
interpreted as HTTP fingerprints observed under the specified experimental
conditions, rather than as universal behavior of each product.

\section{HTTP Request Fingerprints of AI Coding Agents and Assistants}
\label{sec:fingerprints}

\subsection{Experimental Setup}

The controlled endpoint, trial protocol, server-side logging, and analysis
procedure are described in Section~\ref{sec:method}. The following sections
present the observed request fingerprints for AI coding agents and AI assistant
web services.

\noindent\textbf{Empirical findings.} The results below report observed request
sequences, User-Agent strings, header signals, and discovery-file access under
the experimental conditions described in Section~\ref{sec:method}.

New approaches to documentation consumption mean that analytics and maintainers
may now observe only one or two HTTP requests per AI agent session---including a
possible \texttt{/robots.txt} fetch~\cite{palewire2025,sullivan2024}---before
the agent retrieves all content on the page. In this model, analytics no longer
captures how users browse the site, their click patterns, or their scrolling
behavior. Figure~\ref{fig:scroll-events} illustrates these lost client-side events.

\begin{figure}[h]
  \centering
  \setlength{\tabcolsep}{10pt}
  \renewcommand{\arraystretch}{1.35}
  \begin{tabular}{|r|l|}
    \hline
    \textcolor{gray}{1} & \texttt{click} \\
    \hline
    \textcolor{gray}{2} & \texttt{scroll-tag} \\
    \hline
    \textcolor{gray}{3} & \texttt{page\_view} \\
    \hline
    \textcolor{gray}{4} & \texttt{sign\_up} \\
    \hline
    \textcolor{gray}{5} & \texttt{tutorial\_begin} \\
    \hline
    \textcolor{gray}{6} & \texttt{tutorial\_complete} \\
    \hline
    \textcolor{gray}{7} & \texttt{view\_search\_results} \\
    \hline
    \textcolor{gray}{8} & \texttt{login} \\
    \hline
    \textcolor{gray}{9} & \texttt{search} \\
    \hline
    \textcolor{gray}{10} & \texttt{select\_content} \\
    \hline
  \end{tabular}
  \caption{Developer portal analytics event types that become invisible when an
    AI coding agent retrieves documentation in a single server-side request,
    bypassing all client-side instrumentation.}
  \label{fig:scroll-events}
\end{figure}

\subsection{AI Coding Agents}

Table~\ref{tab:agent-http-fingerprints} summarizes the HTTP request
characteristics of nine AI coding agents retrieving developer
documentation. The table shows the HTTP runtime used by each
agent (e.g., Headless Chromium via Playwright, Go \texttt{net/http}, Node.js/Axios,
curl, Node.js/got), the
pre-fetch behavior (on-demand GET, HEAD probe, OpenAPI/Swagger discovery sweep,
sequential multi-page GET), the User-Agent string sent, and the presence of
\texttt{Accept} and \texttt{Sec-Fetch-*} request headers.

\begin{table*}
  \centering
  \caption{HTTP request fingerprints of AI coding agents observed during
           active fetch sessions. Agents are listed alphabetically.
           $\checkmark$~=~header present; \textemdash{}~=~absent.
           \textsuperscript{a}~Full User-Agent strings are available in the
           accompanying session data.}
  \label{tab:agent-http-fingerprints}
  \footnotesize
  \setlength{\tabcolsep}{3pt}
  \begin{tabularx}{\textwidth}{@{} p{1.6cm} p{2.4cm} p{2.6cm} X c c @{}}
    \toprule
    \multirow{2}{*}{\textbf{Agent}} &
    \multirow{2}{*}{\textbf{HTTP Runtime}} &
    \multirow{2}{*}{\textbf{Pre-fetch Behaviour}} &
    \multirow{2}{*}{\textbf{User-Agent\textsuperscript{a}}} &
    \multicolumn{2}{c}{\textbf{Header Signals}} \\
    \cmidrule(lr){5-6}
    & & & & \textbf{Accept} & \textbf{Sec-Fetch-*} \\
    \midrule

    Aider &
    Headless Chromium (Playwright) &
    On-demand GET &
    \texttt{Mozilla/5.0 ([platform]) AppleWebKit/[ver] (KHTML, like Gecko)
            Chrome/145.x Safari/[ver] Aider/0.86.2 +https://aider.chat/} &
    $\checkmark$ & $\checkmark$ \\[2pt]

    Antigravity &
    Go \texttt{net/http} &
    HEAD probe $\rightarrow$ GET &
    \texttt{Go-http-client/2.0} &
    \textemdash{} & \textemdash{} \\[2pt]

    Claude Code &
    Node.js / Axios &
    On-demand GET &
    \texttt{axios/1.8.4} &
    $\checkmark$ & \textemdash{} \\[2pt]

    Cline &
    curl &
    GET + OpenAPI/ Swagger sweep &
    \texttt{curl/8.4.0} &
    $\checkmark$ & \textemdash{} \\[2pt]

    Cursor &
    Node.js / got &
    HEAD probe $\rightarrow$ GET &
    \texttt{got (https://github.com/sindresorhus/got)} &
    \textemdash{} & \textemdash{} \\[2pt]

    Junie &
    curl &
    Sequential multi-page GET &
    \texttt{curl/8.4.0} &
    $\checkmark$ & \textemdash{} \\[2pt]

    OpenCode &
    Headless Chromium (Playwright) &
    On-demand GET &
    \texttt{Mozilla/5.0 ([platform]) AppleWebKit/[ver] (KHTML, like Gecko)
            Chrome/143.x Safari/[ver]} &
    $\checkmark$ & \textemdash{} \\[2pt]

    GitHub Copilot (VS~Code agent mode) &
    Electron / Chromium &
    On-demand GET &
    \texttt{Mozilla/5.0 ([platform]) AppleWebKit/[ver] (KHTML, like Gecko)
            Code/1.111.x Chrome/142.x Electron/39.x Safari/[ver]} &
    $\checkmark$ & $\checkmark$ \\[2pt]

    Windsurf &
    Go / Colly &
    On-demand GET &
    \texttt{colly - https://github.com/gocolly/colly} &
    $\checkmark$ & \textemdash{} \\[2pt]

    \bottomrule
  \end{tabularx}
\end{table*}

An agent that probes \texttt{robots.txt} before fetching
content~\cite{anthropic2024,palewire2025} shows greater compliance with
established web norms. Since an agent---rather than a human---is pulling
information from the page, if website creators have not adapted their
\texttt{robots.txt} to allow or restrict specific content, users may receive
irrelevant or stale data through the agent's interface.

\medskip
\noindent\colorbox{gray!8}{\parbox{\dimexpr\linewidth-2\fboxsep\relax}{%
  \smallskip
  \small\textbf{Observation.}
  Of the agents examined, Aider and OpenCode stand out as the only two that rely
  on Headless Chromium (instantiated via Playwright) as their HTTP runtime. The
  remaining agents employ considerably lighter clients: Go~\texttt{net/http}
  (Antigravity), Node.js/Axios (Claude Code), curl (Cline, Junie), Node.js/got
  (Cursor), and Go/Colly (Windsurf). The practical distinction is non-trivial:
  whereas these lightweight clients issue raw HTTP requests, a Playwright-driven
  headless browser fully evaluates JavaScript prior to resolving the fetch.
  In practice, Aider and OpenCode are better positioned to retrieve
  documentation from client-side-rendered portals, whereas the remaining
  lightweight HTTP clients receive only the static HTML delivered by the server
  and do not execute JavaScript. This approach results in increased resource
  consumption and slower request completion compared to lightweight clients.

  GitHub Copilot (VS~Code agent mode) occupies a somewhat anomalous position in
  this landscape. Although its
  application runtime is built on Electron/Chromium, and its outbound requests
  therefore carry Chromium-style User-Agent strings, HTTP calls originate from
  the extension host layer rather than from full browser navigations. The result
  is a hybrid fingerprint: Chromium identity signals are present, yet the
  requests bypass the complete JavaScript rendering pipeline characteristic of a
  dedicated headless browser session.
  \smallskip
}}
\medskip

\subsection{AI Assistant Web Services}

Table~\ref{tab:assistant-http-fingerprints} presents the same fingerprint
analysis for six AI assistant web services, captured when a URL was shared
inside the chat interface, triggering a server-side fetch by the respective
service backend.

\begin{table*}
  \centering
  \caption{HTTP request fingerprints of six AI assistant web services observed
           when a URL was shared in the chat interface (server-side fetch).
           Assistants are listed alphabetically.
           $\checkmark$~=~header present; \textemdash{}~=~absent.
           \textsuperscript{a}~Full User-Agent strings are available in the
           accompanying session data.
           \textsuperscript{b}~MistralAI issues two distinct requests:
           a lightweight robot probe and a separate browser-mode fetch.}
  \label{tab:assistant-http-fingerprints}
  \footnotesize
  \setlength{\tabcolsep}{3pt}
  \begin{tabularx}{\textwidth}{@{} p{1.7cm} p{2.8cm} p{2.8cm} X p{1.0cm} p{1.6cm} @{}}
    \toprule
    \multirow{2}{*}{\textbf{Assistant}} &
    \multirow{2}{*}{\textbf{HTTP Runtime}} &
    \multirow{2}{*}{\textbf{Pre-fetch Behaviour}} &
    \multirow{2}{*}{\textbf{User-Agent\textsuperscript{a}}} &
    \multicolumn{2}{c}{\textbf{Header Signals}} \\
    \cmidrule(lr){5-6}
    & & & & \textbf{Accept} & \textbf{Sec-Fetch-*} \\
    \midrule

    ChatGPT &
    Custom HTTP (Envoy) &
    On-demand GET &
    \texttt{Mozilla/5.0 AppleWebKit/537.36 (\ldots{};
            ChatGPT-User/1.0; +https://openai.com/bot)} &
    $\checkmark$ & \textemdash{} \\[2pt]

    Claude &
    Custom HTTP client &
    Parallel GET \texttt{/robots.txt} + \texttt{/} (separate IPs) &
    \texttt{Mozilla/5.0 AppleWebKit/537.36 (\ldots{};
            Claude-User/1.0; +Claude-User@anthropic.com)} &
    $\checkmark$ & \textemdash{} \\[2pt]

    Google Gemini &
    Custom HTTP client &
    On-demand GET &
    \texttt{Google} &
    $\checkmark$ & \textemdash{} \\[2pt]

    Google NotebookLM &
    Custom HTTP client &
    On-demand GET &
    \texttt{Google-NotebookLM} &
    \textemdash{} & \textemdash{} \\[2pt]

    MistralAI\textsuperscript{b} &
    Custom HTTP + Headless Chromium &
    \texttt{robots.txt} probe; browser-mode GET \texttt{/} &
    \texttt{MistralAI-User} /
    \texttt{Mozilla/5.0 AppleWebKit/537.36 (\ldots{};
            MistralAI-User/1.0; +https://docs.mistral.ai/robots)} &
    $\checkmark$ & $\checkmark$ \\[2pt]

    Perplexity &
    Custom HTTP client &
    On-demand GET &
    \texttt{Mozilla/5.0 AppleWebKit/537.36 (\ldots{};
            PerplexityUser/1.0; +https://perplexity.ai/perplexity-user)} &
    \textemdash{} & \textemdash{} \\[2pt]

    \bottomrule
  \end{tabularx}
\end{table*}

\begin{table*}
  \centering
  \caption{Trial-level stability of observed HTTP fingerprints. Every tool was
           tested in three independent trials (T1--T3). ``Same'' indicates that
           the attribute was unchanged across all trials. The detailed
           pre-fetch behaviors are reported in
           Tables~\ref{tab:agent-http-fingerprints}
           and~\ref{tab:assistant-http-fingerprints}.}
  \label{tab:trial-stability}
  \footnotesize
  \setlength{\tabcolsep}{3pt}
  \begin{tabularx}{\textwidth}{@{} p{1.7cm} c c c X c c c @{}}
    \toprule
    \textbf{Tool} & \textbf{Trials} & \textbf{User-Agent} &
    \textbf{Headers} & \textbf{Pre-fetch} & \textbf{robots.txt} &
    \textbf{llms.txt} & \textbf{Deviations} \\
    \midrule
    Aider & T1--T3 & Same & Same & Same & No & No & None \\
    Antigravity & T1--T3 & Same & Same & Same & No & No & None \\
    Claude Code & T1--T3 & Same & Same & Same & No & No & None \\
    Cline & T1--T3 & Same & Same & Same & No & No & None \\
    Cursor & T1--T3 & Same & Same & Same & No & No & None \\
    Junie & T1--T3 & Same & Same & Same & No & No & None \\
    OpenCode & T1--T3 & Same & Same & Same & No & No & None \\
    GitHub Copilot (VS~Code agent mode) & T1--T3 & Same & Same & Same & No & No & None \\
    Windsurf & T1--T3 & Same & Same & Same & No & No & None \\
    ChatGPT & T1--T3 & Same & Same & Same & No & No & None \\
    Claude & T1--T3 & Same & Same & Same & Yes & No & None \\
    Google Gemini & T1--T3 & Same & Same & Same & No & No & None \\
    Google NotebookLM & T1--T3 & Same & Same & Same & No & No & None \\
    MistralAI & T1--T3 & Same & Same & Same & Yes & No & None \\
    Perplexity & T1--T3 & Same & Same & Same & No & No & None \\
    \bottomrule
  \end{tabularx}
\end{table*}

Table~\ref{tab:trial-stability} summarizes the stability assessment. No tool
requested \texttt{llms.txt}; access to \texttt{robots.txt} occurred only for
Claude and MistralAI. All recorded User-Agent strings, header sets, and
pre-fetch behaviors were unchanged across the three trials for each tool.
Request counts were also stable: Aider, Claude Code, OpenCode, GitHub Copilot
(VS~Code agent mode), Windsurf, ChatGPT, Google Gemini, Google NotebookLM, and
Perplexity each issued one request per prompt; Antigravity, Cursor, Claude, and
MistralAI each issued two; Junie issued three; and Cline issued four. No
request-count deviations were observed across T1--T3.

\subsection{Comparison of Coding Agents and AI Assistant Services}

Table~\ref{tab:group-comparison} compares the two evaluated tool groups. The
measured attributes, including request counts, summarize the stable patterns
observed across the three trials. The final row is interpretive: it is based on
observed request metadata and does not verify the internal retrieval
architecture of a product.

\begin{table*}
  \centering
  \caption{Comparison of evaluated coding agents and AI assistant services.}
  \label{tab:group-comparison}
  \footnotesize
  \setlength{\tabcolsep}{4pt}
  \begin{tabularx}{\textwidth}{@{} p{3.0cm} X X p{1.8cm} @{}}
    \toprule
    \textbf{Dimension} & \textbf{Coding agents (n=9)} &
    \textbf{AI assistant services (n=6)} & \textbf{Basis} \\
    \midrule
    Server-side requests per prompt &
    1--4 requests per prompt (16 per trial) &
    1--2 requests per prompt (8 per trial) &
    Measured \\
    Pre-fetch behavior &
    Five on-demand GETs; two HEAD-to-GET probes; one OpenAPI/Swagger sweep;
    one sequential multi-page retrieval &
    Four on-demand GETs; two patterns that include a \texttt{robots.txt}
    request and content retrieval &
    Measured \\
    User-Agent identification &
    Usually generic runtime or browser identifiers; Aider and GitHub Copilot
    (VS~Code agent mode) include explicit product identifiers &
    Usually provider or product identifiers; Google Gemini reports only
    \texttt{Google} &
    Measured \\
    Header signals &
    \texttt{Accept} present for 7/9; \texttt{Sec-Fetch-*} present for 2/9 &
    \texttt{Accept} present for 4/6; \texttt{Sec-Fetch-*} present for 1/6 &
    Measured \\
    Discovery-file access &
    No \texttt{robots.txt} or \texttt{llms.txt} requests observed &
    \texttt{robots.txt} requested by Claude and MistralAI; no
    \texttt{llms.txt} requests observed &
    Measured \\
    Apparent retrieval setting &
    Metadata is consistent with a mix of lightweight HTTP clients and
    browser-mediated retrieval &
    Metadata is consistent with provider-side HTTP retrieval, with a
    browser-mode request observed for MistralAI &
    Interpretation \\
    \bottomrule
  \end{tabularx}
\end{table*}

\section{Discussion: Implications for Developer Portals}

The HTTP observations reported in this study establish request-level retrieval
patterns under the experimental conditions described in Section~\ref{sec:method}.
The following implications are interpretive: they connect these observed
patterns to existing literature and practitioner developments in developer
portals. They do not constitute additional empirical measurements of token
usage, feedback-channel adoption, search visibility, developer roles, or UX
outcomes.

\subsection{Implication: Feedback Channels in the Agent Era}

As documentation consumption shifts into AI agents and assistants, the
accessibility of existing support channels becomes a more consequential aspect
of developer experience. High-friction reporting mechanisms, such as complex
registration flows, multi-step forms, or channels reachable only through
traditional web navigation, may reduce the likelihood that users submit issues
or feedback when their primary working environment is an IDE or coding agent.
One emerging response is the creation of public MCP server-based feedback
channels for documentation~\cite{stateofdocs2026}, which allow users to submit
feedback through the same interface through which they consume content and to
summarize the problems they encounter without leaving their working
environment.

\subsection{Illustrative Design Consideration: Tokenomics and Documentation Size}

\noindent\textbf{Illustrative example, not an empirical finding.} This study
did not measure token consumption or documentation sizes across portals. The
following example illustrates why documentation size may matter for
agent-mediated retrieval.
Here, \emph{tokenomics} refers to the practical relationship between
documentation size, available context-window capacity, and retrieval cost.

Broader adoption evidence provides context for this design consideration:
51\% of professional developers report using AI tools daily~\cite{stackoverflow2025},
and GitHub reports that more than 1.1 million public repositories use an LLM SDK,
with 693,867 of these projects created in the preceding 12 months
(+178\% year-over-year)~\cite{github2025}. Cisco DevNet portal users also use
direct links to related OpenAPI documents as input to AI agents or RAG
pipelines~\cite{cisco2024blog}. These external sources motivate the
consideration of documentation size but are not measurements from this study.

The Secure Firewall Management Center REST API Quick Start Guide,
Version~10.0~\cite{cisco2024fw} is reported to contain 193,217 tokens
(717,993 characters). Because this is a single document rather than a
multi-portal sample, it does not establish a general distribution of
documentation sizes or context-window impact. Nevertheless, the example
illustrates how a large document can make available context-window capacity a
practical consideration, depending on an agent's model and configuration.
Documentation-size metadata may help developers and agents decide whether to
use a fresh chat session, apply compression, or use chunked retrieval.
Evaluating these effects requires a dedicated multi-portal study.

\subsection{Implication: Documentation Standards for AI Discovery}

Documentation platforms are introducing machine-readable standards that go beyond
traditional sitemaps and \texttt{robots.txt}. The \texttt{llms.txt} specification
acts as a directory listing all documentation pages with descriptions so that AI
agents know where to find information, while \texttt{skill.md} files provide
structured capability summaries that tell agents what they can accomplish with a
product, what inputs they need, and what constraints apply~\cite{mintlify2024}.
Complementing these discovery mechanisms, Marro et al.~\cite{marro2026} propose
\texttt{agent-\allowbreak{}permissions.json}---a lightweight, \texttt{robots.txt}-style
permission manifest that allows websites to declaratively specify which automated
interactions are allowed, including rate limits, human-in-the-loop requirements,
and preferred API endpoints. Together, these emerging standards form a layered
governance stack: \texttt{llms.txt} guides content discovery, \texttt{skill.md}
describes product capabilities, and \texttt{agent-\allowbreak{}permissions.json} governs how
AI agents may interact with web resources.

\texttt{AGENTS.md}~\cite{agentsmd2026} is becoming the de facto standard for
guiding AI coding agents within software projects. When a developer opens a
project in an AI coding agent, the agent automatically searches for an
\texttt{AGENTS.md} file in the root directory and, if found, incorporates its
instructions and context into every subsequent task and reasoning step.

To improve the developer experience for AI-assisted workflows, Cisco
DevNet introduced \texttt{AGENTS.md} as the default file in the GitHub template
for open-source projects~\cite{ciscodevnet_template}. Under this template, newly created projects include an \texttt{AGENTS.md} file
populated with project-specific content, including direct links to OpenAPI
documentation, DevNet sandboxes~\cite{devnet_sandbox}, and test environments. A well-structured
\texttt{AGENTS.md} improves the experience of working with open-source projects
using AI coding agents by surfacing project structure, dependencies, and relevant
resources directly within the agent's context.

The \emph{Copy for AI} button is one practical feature that documentation
maintainers and Developer Relations teams are already
implementing~\cite{paloalto2024}. It allows copying a page in Markdown or
plain text and pasting it directly into an AI agent prompt, reducing
the prompt token count and delivering all necessary information without HTML tags or
JavaScript. The main trade-off is that it still requires developers to leave
their IDE environment and manually copy the needed content.

\subsection{Implication: Developer Analytics in the AI Era}
\label{sec:analytics}

When AI agents consume developer portal content, portal maintainers typically
observe only one or two requests per session
(Table~\ref{tab:agent-http-fingerprints}), as agents pull information in a
single fetch and guide users to continue interacting within the agent's own
environment. In analytics reports, this appears as a dramatically higher
bounce rate, and user path analysis becomes less useful. All links mentioned in
the agent's response are processed by the agent, but UTM parameters in those
links are stripped. The developer portal is no longer the primary source of
traffic attribution; the AI agent environment is.

Traffic on the developer portal no longer represents the entire audience: a
single AI crawler visit can substitute for many individual user visits, since
end users consume content through coding agents or AI assistants. Traffic
distribution has shifted from organic search and referral sources toward AI
assistant referrals and single-request AI agent sessions.

One practical step is to monitor web analytics for the following AI referral
traffic sources:
\begin{itemize}\setlength{\itemsep}{1pt}
  \item \texttt{labs.perplexity.ai} / referral
  \item \texttt{chatgpt.com} / (none)
  \item \texttt{chatgpt.com} / organic
  \item \texttt{link.edgepilot.com} / referral
  \item \texttt{platform.openai.com} / referral
  \item \texttt{perplexity} / (not set)
  \item \texttt{claude.ai} / referral
  \item \texttt{copilot.microsoft.com} / referral
  \item \texttt{gemini.google.com} / referral
  \item \texttt{chatgpt.com} / (not set)
  \item \texttt{perplexity.ai} / referral
\end{itemize}

\section{Broader Implications for Developer Experience}

\subsection{AI as an Emerging Content Consumer}

Developers are no longer the only personas consuming documentation; AI agents do
so as well. In agent mode, many assistants proceed through documentation and
tasks involving minimal human interaction beyond approving an action. However,
Kumar et al.~\cite{kumar2025} found that developers who actively collaborated
with the agent using an incremental strategy resolved 83\% of real-world issues,
compared to only 38\% with a one-shot delegation approach, indicating that
effective developer--agent collaboration still calls for substantial human
involvement. Kuo et al.~\cite{kuo2026} found that developers are most
receptive to proactive AI assistance at natural workflow boundaries, while
mid-task interventions are dismissed 62\% of the time, suggesting that
documentation and suggestions surfaced by AI agents must correspond to the
developer's cognitive context.

One hypothesis is that, as more tasks are completed inside AI assistant
interfaces or IDEs, portal teams may need to reassess the balance between visual
UI optimization and machine-readable content. Answer Engine Optimization (AEO),
Generative Engine Optimization (GEO), AI-native features, and MCP servers are
possible directions, but their effectiveness was not evaluated in this study.
Aggarwal et al.~\cite{aggarwal2023} report visibility effects for generative
engine optimization in their setting; whether those effects apply to developer
documentation requires separate evaluation.

The concept of who consumes developer content is expanding beyond human individuals.
Combemale~\cite{combemale2025} notes that the developer community now includes
``scientists, domain experts, and even non-programmers---often referred to as
`citizen developers'---increasingly engaging in programming to advance their
fields or solve domain-specific problems.'' AI agents represent the next
frontier of this expanding audience, acting as autonomous consumers of
documentation on behalf of human developers.

\subsection{Hypothesis: Emerging Roles in Developer Relations}

In developer relations, professionals are currently responsible for community
communication and gathering developer feedback. The traffic patterns observed
here may motivate new specialist roles, such as Agentic Engine Optimization
engineers and AI DevRel managers, in developer relations organizations. Such
roles could focus on monitoring which content is consumed by AI
coding agents and assistants and on ensuring that content remains visible,
accurate, and well-structured within those environments.

\subsection{Hypothesis: UX and UI Transformation}

The UX and design of documentation and developer guides may become less central
to some interactions as activity migrates to IDEs and AI agents. Developers and
engineers may increasingly rely on AI tools to retrieve and synthesise
information that they would otherwise access through traditional site navigation.
AI browsers such as Perplexity Comet, Google Chrome with Gemini, and
ChatGPT's browsing mode, as well as AI crawlers, are changing how developers
consume content. Developers interact with developer portals via free-text fields
embedded in AI browsers or agents; as a result, the observed single-request
access patterns suggest that traditional A/B testing and UI interaction
measurement approaches may require fundamental rethinking in
AI-agent-dominated traffic environments. Users do not need to notice individual
UI elements; AI browsers can read HTML text, including button alt text. In such
workflows, content structure and machine-readability may become more important
than button colour and placement.

\section{Limitations}

Our study has a number of limitations that should be kept in mind when
interpreting the results. First, we restricted our agent corpus to tools that
were freely available, open source, or offered a free/trial tier at the time of
data collection; proprietary agents without public access were excluded. Whether
the behavioral signatures we report generalize to closed commercial deployments
therefore remains an open question. Second, several open-source agents relied on
third-party LLM providers or external web search services to handle
documentation retrieval. In those cases, the HTTP traffic we captured reflects
the joint behavior of the agent and its upstream dependencies, making it
difficult to attribute individual observations to the agent alone. Third, our
measurements are point-in-time snapshots tied to specific software versions.
Given the pace at which AI coding agents are being updated, characteristics such
as User-Agent headers, pre-fetch strategies, and header patterns may shift
substantially across releases. Longitudinal monitoring of agent HTTP behavior,
as well as expansion of the corpus to agents available only under enterprise
licensing, are natural directions for future work. In addition, the measurements
were collected against a single documentation endpoint, so agent behavior may
differ for portals with alternative rendering strategies, authentication
requirements, or anti-bot protections.

\section{Conclusion}

The proliferation of AI coding agents and assistants is fundamentally reshaping
how developers discover, consume, and evaluate documentation. This paper
has examined the transformation across three interconnected dimensions:
documentation accessibility, content analytics, and feedback systems.

Our empirical analysis of HTTP request characteristics from nine AI coding
agents (Table~\ref{tab:agent-http-fingerprints}) and six AI assistants
(Table~\ref{tab:assistant-http-fingerprints}) reveals identifiable behavioral
signatures---runtime environments, pre-fetch strategies, User-Agent strings, and
header patterns---that developer portal maintainers can leverage to detect and
classify AI-driven traffic directly from web server logs. A defining feature of
AI agent access is the compression of multi-page navigation into one or two
requests, which renders traditional engagement metrics such as session depth,
time-on-page, click-path, and bounce rate unreliable indicators of actual
documentation consumption.

The tokenomics of documentation is an illustrative, underexplored design
consideration for developer experience. As documentation grows relative to an
agent's available context window, developers and agents may need to consider
chunking, retrieval, and context management. The reported 193,217-token Cisco
Firewall Management Center REST API guide provides one example, rather than
evidence of a general trend. Surfacing token counts alongside documentation
entry points may support context management; evaluating its effects on task
completion, cost, and retrieval requires a dedicated multi-portal study.

The emerging standards ecosystem---\texttt{llms.txt} for content discovery,
\texttt{skill.md} for capability description, and \texttt{agent-\allowbreak{}permissions.json}
for interaction governance---provides portal teams with structured mechanisms to
restore visibility and control over how content is discovered and consumed by
autonomous agents. Pairing these standards with MCP server--based feedback
channels and analytics instrumented for AI referral sources enables a more
complete picture of developer engagement in the agent era.

Broader implications for developer relations are hypotheses for future
research. If documentation consumption continues to migrate to AI agent
environments and IDEs, portal maintainers may need to examine the balance
between visual UI optimization, AEO/GEO practices, and AI-native content
structures. As Aggarwal et al.~\cite{aggarwal2023} show, such changes can boost
source visibility by up to 40\% in their setting; this study did not evaluate
whether that result applies to developer portals. These shifts
could also motivate specialist roles, such as Agentic Engine Optimization
engineers and AI DevRel managers. Their emergence and effectiveness require
separate study.

Potential design considerations include auditing
\texttt{robots.txt} files to verify that AI crawlers can reach the intended
developer-oriented content, publishing \texttt{llms.txt} and stable OpenAPI
specification endpoints, and surfacing token counts on large documentation pages
where appropriate. These considerations are consistent with the traffic
patterns observed in this study but require independent evaluation. Instrumenting
analytics to recognize the AI referral sources identified in
\S\ref{sec:analytics} may provide visibility into agent-driven traffic that
standard reporting does not capture.

Future work should address standardized telemetry formats for AI-driven
documentation traffic, longitudinal studies measuring how GEO-optimized content
affects agent task completion rates, and the design of machine-readable feedback
protocols that enable AI agents to report documentation quality signals on behalf
of users.

\section*{Supplementary Replication Evidence: Illustrative Request Traces}

Listings~\ref{lst:opencode-trace} and~\ref{lst:mistral-trace} present two
anonymized request traces captured on 2026-03-10, one of the trial dates within
the February--March 2026 measurement period. They illustrate the request
metadata used in the fingerprint analysis; the full anonymized logs are
available in the companion repository described in the Data Availability
statement. Client IP addresses and empty headers are omitted.

\begin{lstlisting}[style=requesttrace,
caption={Anonymized OpenCode request trace.},
label={lst:opencode-trace}]
{
  "agent": "OpenCode",
  "timestamp": "2026-03-10TXX:XX:XX+XX:XX",
  "method": "GET",
  "url": "/",
  "httpVersion": "HTTP/1.1",
  "statusCode": 200,
  "bodyBytesSent": 12546,
  "headers": {
    "host": "[second-level domain].dev",
    "userAgent": "Mozilla/5.0 ([platform]) AppleWebKit/[ver] (KHTML, like Gecko) Chrome/143.x Safari/[ver]",
    "accept": "text/html;q=1.0, application/xhtml+xml;q=0.9, text/plain;q=0.8, text/markdown;q=0.7, */*;q=0.1",
    "acceptEncoding": "gzip, deflate, br, zstd",
    "acceptLanguage": "en-US,en;q=0.9"
  }
}
\end{lstlisting}

\begin{lstlisting}[style=requesttrace,
caption={Anonymized MistralAI web request trace.},
label={lst:mistral-trace}]
{
  "agent": "MistralAI (web)",
  "timestamp": "2026-03-10TXX:XX:XX+XX:XX",
  "method": "GET",
  "url": "/",
  "httpVersion": "HTTP/1.1",
  "statusCode": 200,
  "bodyBytesSent": 12546,
  "headers": {
    "host": "[second-level domain].dev",
    "upgradeInsecureRequests": "1",
    "userAgent": "Mozilla/5.0 AppleWebKit/537.36 (KHTML, like Gecko; compatible; MistralAI-User/1.0; +https://docs.mistral.ai/robots)",
    "accept": "text/html,application/xhtml+xml,application/xml;q=0.9,image/avif,image/webp,image/apng,*/*;q=0.8,application/signed-exchange;v=b3;q=0.7",
    "acceptEncoding": "gzip, deflate, br, zstd",
    "acceptLanguage": "en-US,en;q=0.9",
    "secFetchSite": "none",
    "secFetchMode": "navigate",
    "secFetchDest": "document",
    "secFetchUser": "?1"
  }
}
\end{lstlisting}

\paragraph*{How to Cite.}
A short, peer-reviewed version of this work was published in the ACM Digital
Library. Please cite it as the primary reference:

\begin{quote}\small
Oleksii Borysenko. 2026. Documentation Retrieval by AI Coding Agents: An
Empirical Study of HTTP Behavioral Signatures. In \emph{Proceedings of the 1st
International Workshop on Agentic AI for Next-Generation Software Development
(AgenticDev '26)}, Munich, Germany. ACM, New York, NY, USA, 168--169.
\url{https://doi.org/10.1145/3843282.3844445}
\end{quote}

\noindent When referring to the extended method, results, discussion, or
replication material that appears only in this version, please additionally
cite:

\begin{quote}\small
Oleksii Borysenko. 2026. Developer Experience with AI Coding Agents: HTTP
Behavioral Signatures in Documentation Portals. arXiv:2604.02544 [cs.SE].
\url{https://arxiv.org/abs/2604.02544}
\end{quote}

\paragraph*{Data Availability.}
The HTTP fingerprint data reported in this work were collected from a
purpose-built developer documentation endpoint operated by the authors. The
endpoint served developer-oriented content and was configured with both
\texttt{robots.txt} and \texttt{llms.txt} files to approximate the conditions
of a production documentation portal. The anonymized raw HTTP logs are
available in the companion repository:
\url{https://github.com/oborys/AI-Agents-HTTP-level-behaviour}. Summary
fingerprint data are given in Tables~\ref{tab:agent-http-fingerprints}
and~\ref{tab:assistant-http-fingerprints}. Any IP addresses or personally
identifiable information incidentally recorded during data collection are
redacted from shared data, consistent with applicable privacy requirements.

\bibliographystyle{unsrt}
\bibliography{references}

\end{document}